\pdfoutput=1   % arXiv: build with pdfLaTeX (PNG figures have no EPS BoundingBox)
\documentclass[
  aps,
  pre,
  reprint,
  amsmath,amssymb,
  superscriptaddress,
  floatfix,
  longbibliography,
  nofootinbib
]{revtex4-2}

\usepackage{graphicx}
\usepackage{mathtools}
\usepackage{subcaption}
\usepackage[noend]{algpseudocode}
\usepackage{placeins}   % \FloatBarrier keeps floats inside their section
\usepackage{hyperref}
\hypersetup{colorlinks=true,citecolor=blue,linkcolor=blue,urlcolor=blue}

\newcounter{algorithm}
\renewcommand{\thealgorithm}{\arabic{algorithm}}
\makeatletter
\newenvironment{algorithm}[1][]{%
  \par\addvspace{8\p@}%
  \refstepcounter{algorithm}%
  \hrule\@height.8\p@\vskip3\p@
  \renewcommand{\caption}[1]{%
    \noindent\textbf{Algorithm \thealgorithm:} ##1\par
    \vskip2\p@\hrule\@height.4\p@\vskip4\p@}%
}{%
  \vskip3\p@\hrule\@height.8\p@\par\addvspace{8\p@}%
}
\makeatother

\begin{document}

\title{Algorithmic overlaps in the Baxter--Wu model:
cluster dynamics under Novotny--Evertz updates}

\author{Ian Pil\'e}
\thanks{Corresponding author}
\email{pileyan@gmail.com}
\affiliation{HSE University, Moscow 101000, Russia}

\author{Lev Shchur}
\affiliation{HSE University, Moscow 101000, Russia}

\date{\today}

\begin{abstract}
We study the spatial overlap of successive spin configurations generated by Markov chain Monte Carlo simulations of the Baxter--Wu model. Using the Novotny--Evertz sublattice-freezing single-cluster update, we track the mean and variance of the algorithmic overlap across the critical region. We show that, even in this three-spin model, the overlap acts as an algorithmic observable that follows the thermodynamics of the transition: the single-cluster overlap mean behaves like an order parameter, dropping from a finite ordered-phase plateau toward zero across $T_c$. The overlap does not diverge at criticality, instead it remains finite and its finite-size value decays as a clean power law, $U_2(T_c)\sim L^{-\psi}$, over eleven sizes with an exponent $\psi{=}0.378(4)$ smaller than the value $\approx0.42$ found for the Ising and Potts models under standard Fortuin--Kasteleyn cluster dynamics, indicating that it reflects the Novotny--Evertz sublattice-freezing dynamics rather than any static property of the model.
\end{abstract}


\maketitle

% \begin{highlights}
% \item The algorithmic configuration overlap tracks the Baxter--Wu transition.
% \item The single-cluster overlap mean behaves like an order parameter but stays finite---it does not diverge---at $T_c$.
% \item Its finite-size exponent $\psi^{(\mathrm{W})}{=}0.378(4)$ is smaller than the Ising and Potts values and reflects the Novotny--Evertz cluster dynamics.
% \end{highlights}

\section{Introduction}\label{sec:intro}

The Baxter--Wu model is a two-dimensional Ising system on the triangular lattice whose spins interact in triples over the elementary plaquettes~\cite{wood1972,baxter1973}. It is exactly solvable: Baxter and Wu obtained the free energy via a mapping to a coloring problem, finding a continuous transition with $\alpha{=}2/3$ at the self-dual point $k_BT_c/J{=}2/\ln(1+\sqrt2)\approx2.269185$, numerically equal to the square-lattice Ising value~\cite{baxter1973,baxterwu1974,baxter1982book}. The four-fold degeneracy of its ground state places the model in the four-state Potts universality class~\cite{domany1978}, with which it shares the leading exponents ($\beta{=}1/12$, $\nu{=}2/3$) and central charge $c{=}1$~\cite{alcaraz1997}. Its critical behavior has been studied by both local~\cite{costa2004} and cluster~\cite{vasilopoulos2026} methods, so the model is a convenient testing ground for the present work.

The triangular lattice partitions into three interpenetrating sublattices, with every elementary triangle carrying one site of each.
This structure is behind the four-fold ground-state degeneracy, and it also shapes the cluster update itself. A three-spin interaction has no natural bond representation, so the algorithm of Novotny and Evertz~\cite{novotny1989,evertz2003} freezes one randomly chosen sublattice at each step. This reduces the plaquette coupling to an effective pairwise Ising model on the remaining two sublattices, which form a diluted honeycomb lattice. The standard Fortuin--Kasteleyn~\cite{fortuin1972} construction and the Wolff~\cite{wolff1989} single-cluster update then apply directly. Deng \textit{et al.}~\cite{deng2010} later extended this idea to self-dual generalizations. A recent percolation analysis~\cite{vasilopoulos2026} confirmed that these clusters percolate exactly at $T_c$ and reproduce the thermal exponents, while making clear that they are \emph{not} Fortuin--Kasteleyn clusters of the model: the same spin configuration yields three different decompositions depending on which sublattice is frozen, so the clusters are objects of the algorithm rather than a static representation of the Boltzmann weight.

Quantities internal to a Monte Carlo algorithm can themselves behave as thermodynamic functions. The clearest example is the acceptance rate of local updates, which is a function of the internal energy~\cite{lev2019}; related ideas appear in generalized-ensemble sampling~\cite{machta2010} and in the persistence probability of local dynamics~\cite{derrida1994persistence}. The configuration overlap between successive states is another quantity of this kind. For the Ising and $q$-state Potts models it has been shown~\cite{pile2026overlaps} that the single-cluster overlap mean acts as an algorithmic order parameter, while its variance develops a critical peak, with a finite-size-scaling exponent of the mean that takes a common value $\approx0.42$ across the Ising, three- and four-state Potts models---suggesting it reflects cluster geometry rather than any static exponent.

In this work we test whether this algorithmic universality survives in a model with multi-spin interactions. We measure the mean and variance of the algorithmic overlap versus temperature, energy, and system size for the Baxter--Wu model under Novotny--Evertz sublattice-freezing cluster dynamics.
% Fitting eleven system sizes at $T_c$ with independent-replica statistics, we find a clean single-cluster power law $U_2^{(\mathrm{W})}(T_c)\sim L^{-\psi}$ with $\psi{=}0.378(4)$---a well-defined algorithmic exponent, smaller than the value $\approx0.42$ obtained for the Ising and Potts models. The overlap itself stays finite at criticality, so this is a decay exponent of the finite-size value, not a divergence. Since the Baxter--Wu clusters have no static FK counterpart, this exponent cannot descend from a static cluster geometry; it is a property of the freezing dynamics itself, and its shift relative to the FK-based value shows that the algorithmic overlap exponent is sensitive to how the clusters are constructed.

\section{Model}\label{sec:model}

The Baxter--Wu model has $N{=}L^2$ Ising spins $s_i{=}\pm1$ on a triangular lattice with energy
\begin{equation}
\mathcal{H}_{\mathrm{BW}} {=} - J \sum_{\langle ijk\rangle\in\triangle} s_i s_j s_k ,
\label{eq:H_BW}
\end{equation}
summed over elementary triangles ($J\equiv1$).  For comparison we simulate the four-state Potts model on the square lattice,
\begin{equation}
\mathcal{H}_{\mathrm{Potts}}{=}-J\sum_{\langle ij\rangle}\delta_{s_i,s_j} ,
\label{eq:H_Potts}
\end{equation}
with $s_i\in\{1,\dots,4\}$ and $T_c^{(4)}{=}1/\ln3\approx0.910239$~\cite{pottsTc}, which shares the Baxter--Wu universality class.

\begin{figure}[!htb]
  \centering
  \includegraphics[width=\columnwidth]{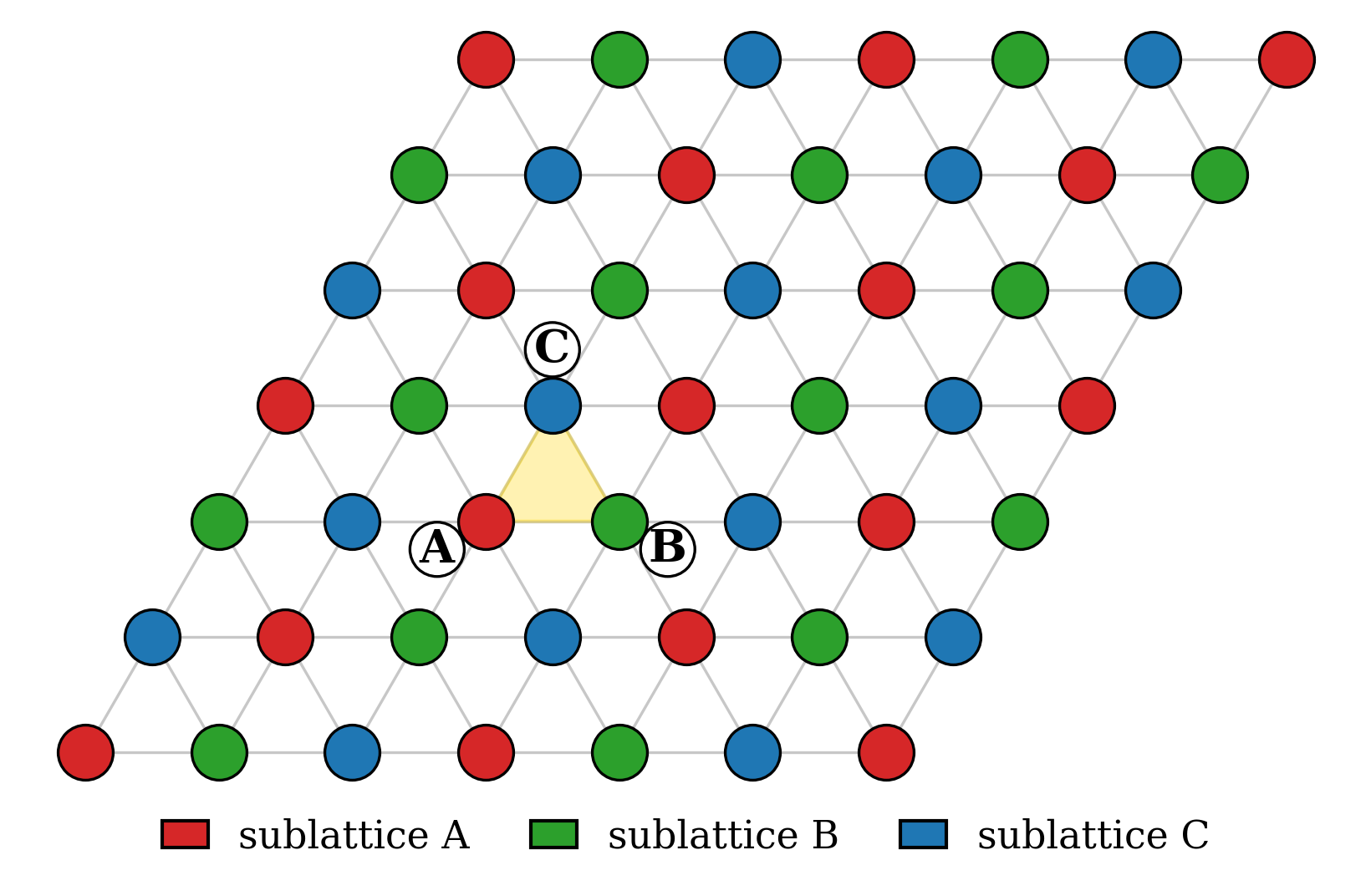}
  \caption{Three-sublattice coloring of the triangular lattice ($A$ red, $B$ green, $C$ blue); every elementary triangle (gold) has one site of each. The unique three-coloring closes on the torus only when $L\bmod3{=}0$. This structure underlies both the four-fold degeneracy and the Novotny--Evertz cluster update.}
  \label{fig:sublattices}
\end{figure}

The triangular lattice splits into three sublattices $A,B,C$, one per triangle corner (Fig.~\ref{fig:sublattices}). The triangular lattice is uniquely three-colorable, so along each lattice axis the colors repeat in the periodic sequence $A,B,C,A,B,C,\dots$~\cite{deng2010}; on a torus this coloring closes consistently---and the three ferrimagnetic ground states fit on the lattice~\cite{vasilopoulos2026}---only when both linear dimensions are multiples of three~\cite{Shchur}. We therefore take $L\bmod3{=}0$ throughout, as required for the Novotny--Evertz construction; all our sizes ($L$ up to $1023$, eleven values in the finite-size-scaling set) satisfy it.

\section{Algorithms and measurement details}\label{sec:algo}

\subsection{Update algorithms}

For the four-state Potts model we use the standard Wolff single-cluster update~\cite{wolff1989}: a seed site is picked at random, and the cluster grows by adding each same-state nearest neighbor of an already-included site with the bond probability
\begin{equation}
p_{\mathrm{bond}} {=} 1 - e^{-\beta J},
\label{eq:bond_probability}
\end{equation}
after which the whole cluster is reassigned to a new state drawn uniformly from the four.

\paragraph*{Wolff--Novotny--Evertz single-cluster update.}
As it was already said, Baxter-Wu model does not admit direct Fortuin-Kasteleyn representation necessary for the Wolff algorithm. The Novotny--Evertz prescription~\cite{novotny1989,evertz2003,vasilopoulos2026} removes this obstacle by freezing one sublattice at each step. At the start of each update step one of the three sublattices---say $C$---is selected uniformly at random and its spins are held fixed. Every elementary triangle contains exactly one $A$, one $B$, and one $C$ site, so each triangle now couples a single pair of \emph{active} spins, one on $A$ and one on $B$, weighted by the frozen $C$ spin sitting at its third corner. The active spins live on the two remaining sublattices, which together form a honeycomb lattice of $2N/3$ sites.

Each active bond $(i,j)$ of this honeycomb lattice belongs to exactly two elementary triangles, one on each side, whose third (frozen) corners carry spins $\sigma_{\perp,1}$ and $\sigma_{\perp,2}$. Summing the two triangle terms that contain the bond gives an effective pairwise Ising coupling between the active spins $s_i$ and $s_j$,
\begin{equation}
J'_{ij} {=} J\,(\sigma_{\perp,1} + \sigma_{\perp,2}) \;\in\; \{-2J,\,0,\,+2J\},
\label{eq:Jeff}
\end{equation}
so the frozen configuration turns the Baxter--Wu model into spatially inhomogeneous nearest-neighbor Ising model.
A bond is ferromagnetic when $J'_{ij}{=}+2J$ (both frozen corner spins are $+1$), antiferromagnetic when $J'_{ij}{=}-2J$ (both $-1$), and absent when $J'_{ij}{=}0$ (the two frozen spins disagree), in which case the two triangle contributions cancel and the bond carries no coupling.

The Fortuin--Kasteleyn rule is now applied to this effective Ising model. A bond $(i,j)$ is eligible to be occupied only if it is satisfied--that is, if $\mathrm{sgn}(J'_{ij})\,s_i s_j {=} +1$, meaning the two active spins are aligned for a ferromagnetic bond or anti-aligned for an antiferromagnetic one--and it is then occupied with probability
\begin{equation}
p_{ij} {=} 1 - e^{-2\beta |J'_{ij}|} {=} 1 - e^{-4\beta J},
\label{eq:wne_bond}
\end{equation}
Unsatisfied and absent bonds are never occupied. We grow a single cluster from a randomly chosen active seed: starting from the seed, we examine its active bonds, occupy each eligible one with probability~(\ref{eq:wne_bond}), add the newly connected active sites to a queue, and repeat until the queue is empty (a breadth-first flood fill). The completed cluster is then flipped, $s_k \to -s_k$ for every active site $k$ in it. Because the underlying effective model is a genuine two-state Ising system, this flip satisfies detailed balance with respect to the Boltzmann weight at fixed frozen sublattice; averaging over the random choice of frozen sublattice, performed independently at each step, preserves detailed balance for the full Baxter--Wu Hamiltonian and restores ergodicity (the frozen sublattice is itself updated whenever it is one of the two active sublattices on a later step).

\begin{algorithm}[!t]
\caption{One Wolff--Novotny--Evertz single-cluster update}
\label{alg:wne}
\begin{algorithmic}[1]
\State pick a sublattice $F\in\{A,B,C\}$ uniformly at random, freeze its spins
\State let $\mathcal{V}$ be the active sites (the two non-frozen sublattices)
\State pick a seed $v_0\in\mathcal{V}$ uniformly at random
\State initialize cluster $\mathcal{C}\gets\{v_0\}$ and queue $Q\gets\{v_0\}$
\While{$Q$ not empty}
  \State pop site $i$ from $Q$
  \For{each active neighbor $j$ of $i$ with $j\notin\mathcal{C}$}
    \State compute $J'_{ij}$ from the two frozen corner spins, Eq.~(\ref{eq:Jeff})
    \If{$J'_{ij}\neq0$ \textbf{and} $\mathrm{sgn}(J'_{ij})\,s_i s_j = +1$}
      \State with probability $p_{ij}$ of Eq.~(\ref{eq:wne_bond}): add $j$ to $\mathcal{C}$ and $Q$
    \EndIf
  \EndFor
\EndWhile
\State flip every spin in $\mathcal{C}$: $s_k\gets-s_k$
\end{algorithmic}
\end{algorithm}

The resulting clusters are objects of the algorithm rather than a static representation of the Boltzmann weight, since the same spin configuration gives three different decompositions depending on which sublattice is frozen~\cite{vasilopoulos2026}. Throughout, one step of the cluster dynamics denotes a single execution of Algorithm~\ref{alg:wne}, and the overlap is measured after each step.

\subsection{Observables}

For the single-cluster update we measure the geometric overlap of the spins flipped by two successive Wolff clusters,
\begin{equation}
U^{(\mathrm{W})}_{n} {=} \frac{1}{N}\,\big|C^{(t)}\cap C^{(t+n)}\big| ,
\label{eq:wolff_overlap}
\end{equation}
where $C^{(t)}$ is the set of sites in the cluster grown at step $t$ and $n$ is the step separation. We record the mean $U^{(\mathrm{W})}_n$ and the variance $\mathrm{Var}(U^{(\mathrm{W})}_n)$ over the run, for separations $n{=}1,\dots,6$. As a thermodynamic reference we compute the heat capacity $C{=}N(\langle\epsilon^2\rangle-\langle\epsilon\rangle^2)/T^2$, with $\epsilon{=}E/N$ the energy per spin.

\subsection{Simulation protocol}

We simulated eleven Baxter--Wu sizes, $L{=}129$, $180$, $255$, $360$, $513$, $591$, $648$, $690$, $768$, $891$, and $1023$ together with a square-lattice four-state Potts system of linear size $L{=}1024$ for comparison. Two kinds of run were performed. For the temperature dependence of the mean and variance (Figs.~\ref{fig:cluster-mean-T}--\ref{fig:bw-vs-potts-loglog}) we scanned $T$ across the critical region, discarding $2\times10^{5}$ cluster steps for equilibration and then averaging over $10^{6}$ measured steps at each temperature (rising to $1.5\times10^{6}$ for the largest sizes). For the finite-size-scaling estimate of $\psi^{(\mathrm{W})}$ we ran longer simulations fixed at $T_c$, described next.

The exponent $\psi^{(\mathrm{W})}$ is the central quantity of the paper, and near $T_c$ the single-cluster overlap is strongly autocorrelated, so a single long chain gives a misleadingly small error bar. We therefore ran, for each $L$, a set of independent replicas started from different random configurations and thermalized separately: between $13$ and $48$ replicas per size, about $400$ in total, with more replicas allocated to the larger sizes where autocorrelation is worst. Each replica used $3\times10^{5}$ equilibration steps for $L\le360$, $6\times10^{5}$ for $L\le690$, and $10^{6}$ for the three largest sizes, followed by up to $4\times10^{5}$ recorded measurements.

The integrated autocorrelation time $\tau_{\mathrm{int}}$ of the overlap was measured directly from each chain. We formed the normalized autocorrelation function of the two-step overlap, summed it with the automatic-windowing procedure of Ref.~\cite{sokal}, and truncated the window $W$ self-consistently at the first $W\ge6\,\tau_{\mathrm{int}}(W)$. A short pilot run at each $L$ gave an initial $\tau_{\mathrm{int}}$, and measurements were then recorded once every $2\,\tau_{\mathrm{int}}$ steps so that the stored samples were close to independent. The residual autocorrelation time of the thinned series was small at most sizes, $\tau_{\mathrm{int}}\lesssim3$, but grew towards the largest lattices, reaching $\tau_{\mathrm{int}}\approx25$ at $L{=}690$, $891$, and $1023$. The number of effectively independent samples, $n_{\mathrm{eff}}=R\,N_{\mathrm{meas}}/2\tau_{\mathrm{int}}$, was kept above $10^{6}$ for every size.

The overlap at $T_c$ and its uncertainty were obtained by a leave-one-out jackknife over the replica means, which treats each replica as one independent measurement and is insensitive to residual within-chain correlation. As a cross-check we also computed a binning error inside each chain, coarsening the bins until the estimate stopped changing, and pooled it across replicas. The two error estimates agreed at every size, confirming that autocorrelation and thermalization were under control. The exponent followed from a two-parameter weighted fit $U_2^{(\mathrm{W})}(T_c)=A\,L^{-\psi^{(\mathrm{W})}}$ to the eleven points, and we checked its stability with an $L_{\mathrm{min}}$ scan in which the smallest sizes are removed one at a time.

\section{Results}\label{sec:results}

\subsection{Cluster overlap as an algorithmic order parameter}

\begin{figure*}[!t]
  \centering
  \begin{subfigure}{0.32\textwidth}
    \centering\includegraphics[width=\linewidth]{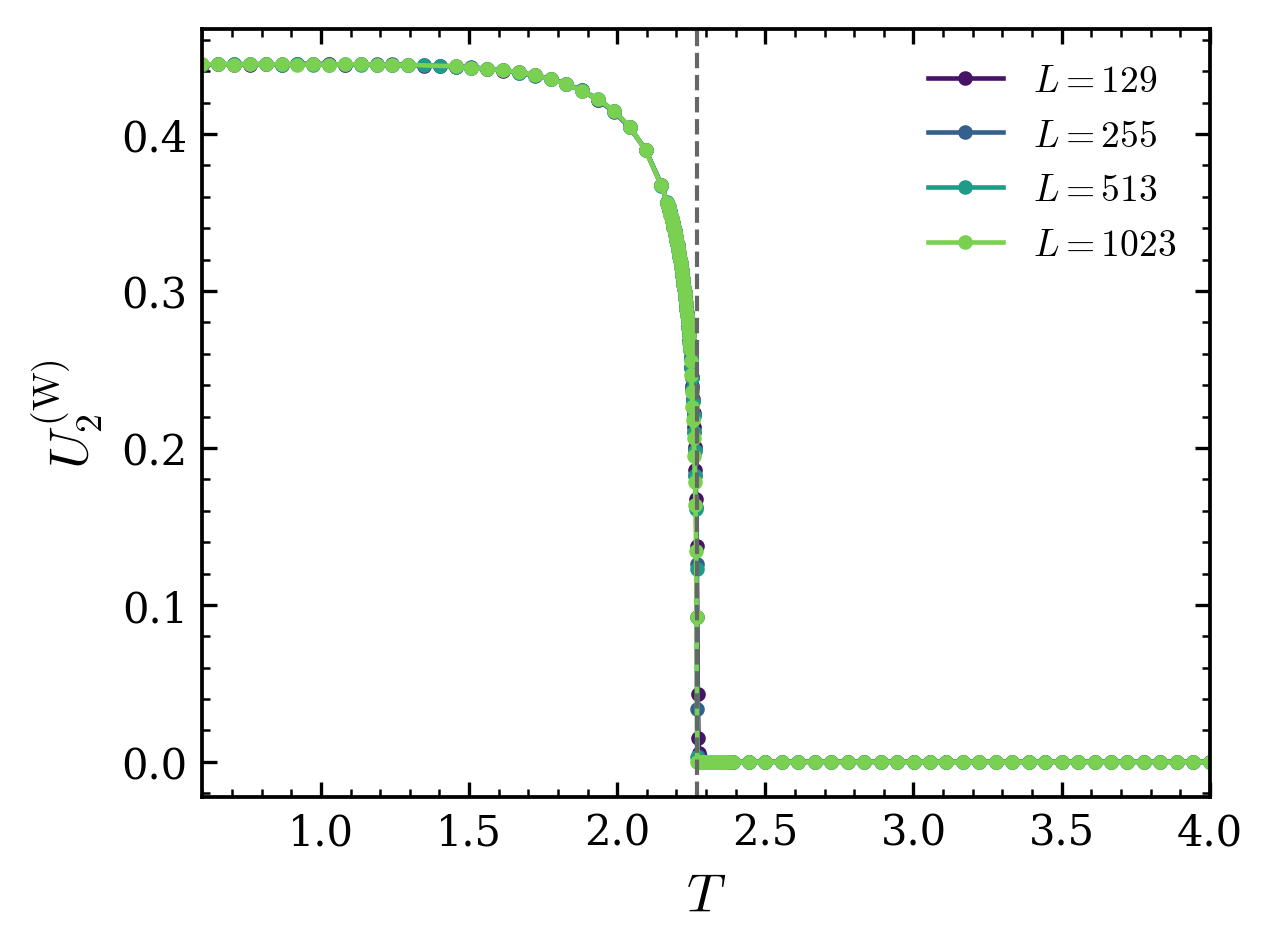}
    \caption{}\label{fig:cluster-mean-T}
  \end{subfigure}\hfill
  \begin{subfigure}{0.32\textwidth}
    \centering\includegraphics[width=\linewidth]{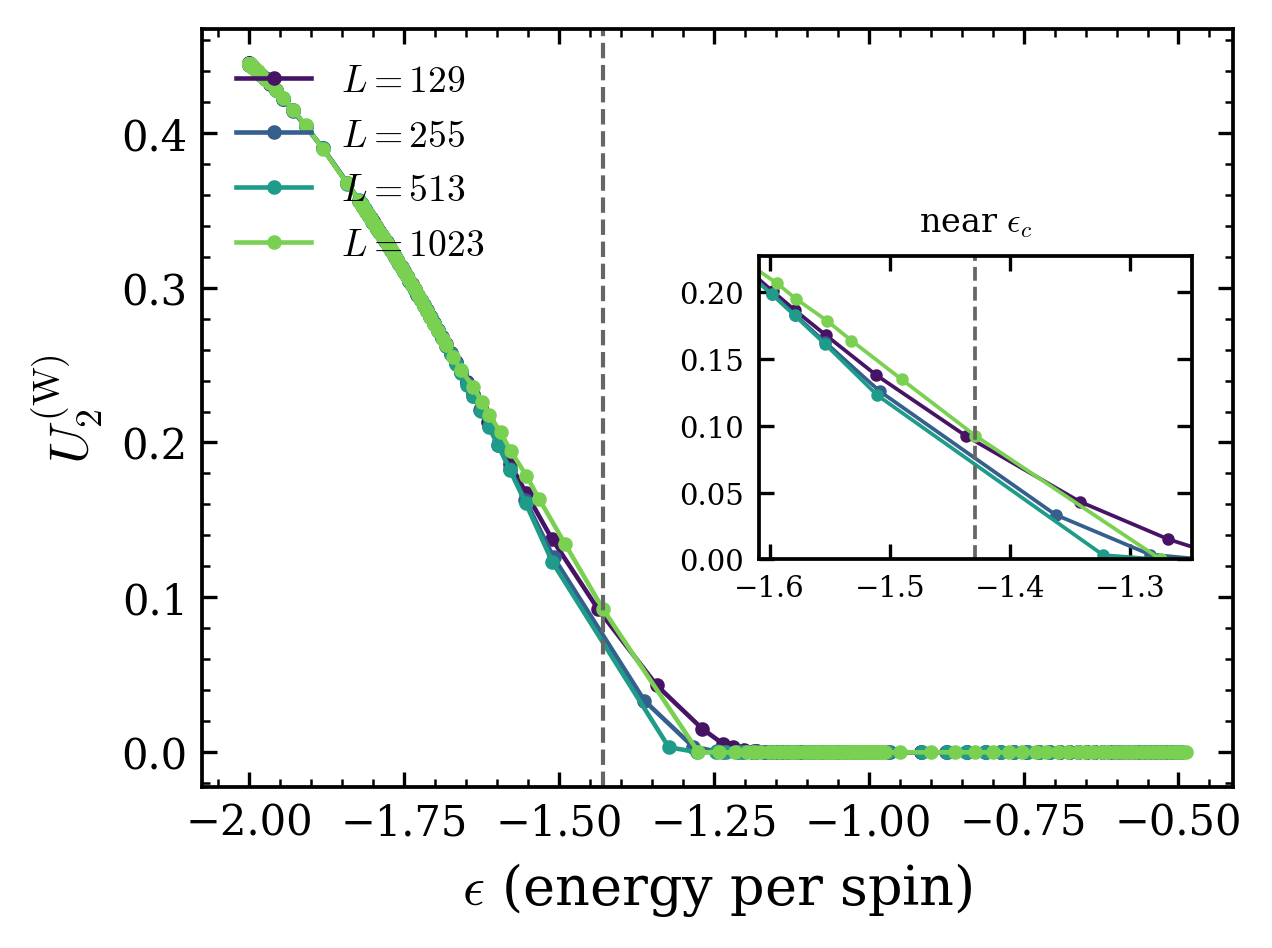}
    \caption{}\label{fig:cluster-mean-E}
  \end{subfigure}\hfill
  \begin{subfigure}{0.32\textwidth}
    \centering\includegraphics[width=\linewidth]{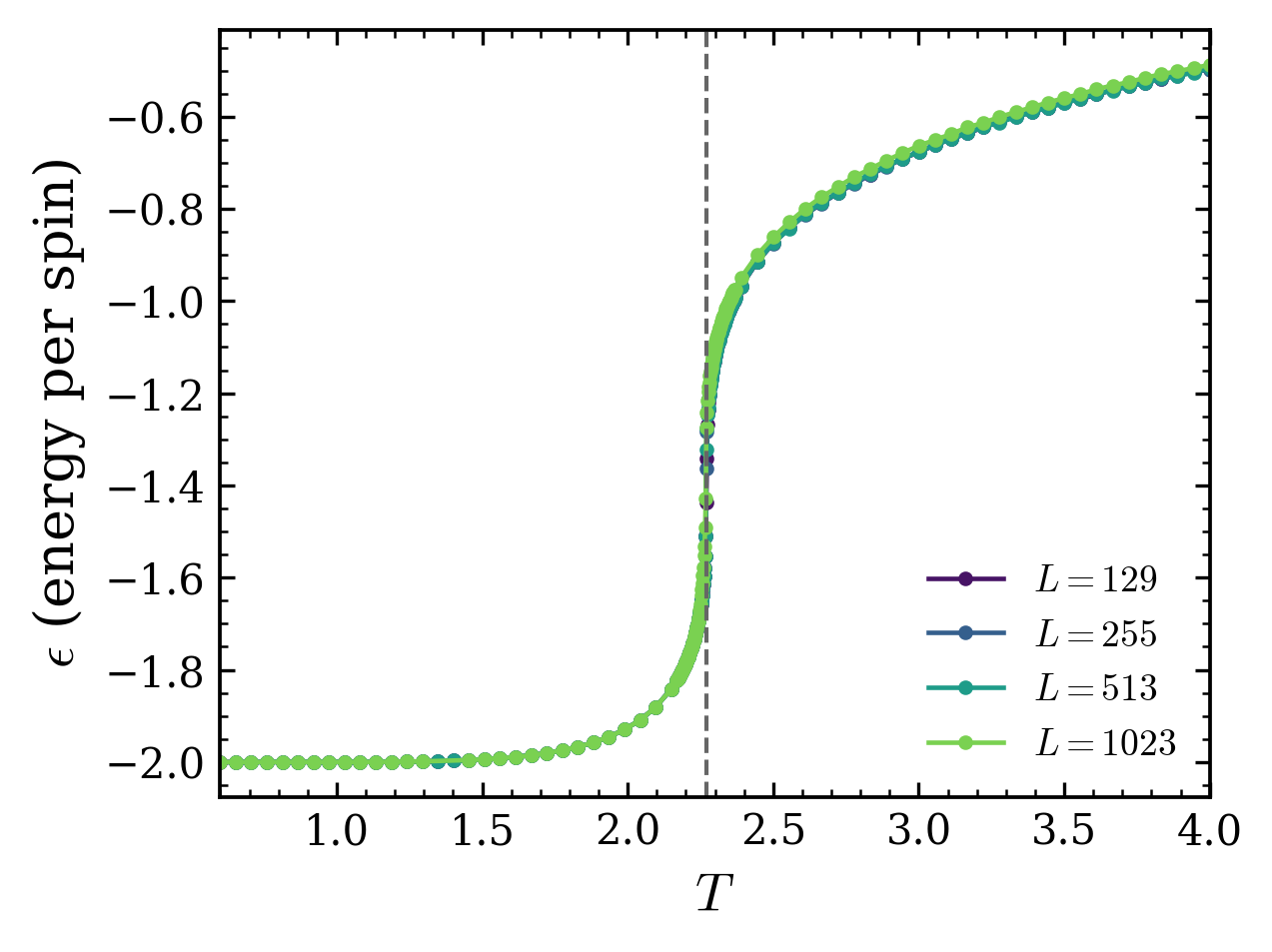}
    \caption{}\label{fig:energy-T}
  \end{subfigure}
  \caption{Baxter--Wu, single-cluster updates. (\subref{fig:cluster-mean-T})~Mean overlap $U_2^{(\mathrm{W})}$ versus $T$ for $L{=}129$--$1023$. The ordered-phase plateau at $\approx0.44$ reflects the $2N/3$ active spins [Eq.~(\ref{eq:plateau})]. (\subref{fig:cluster-mean-E})~Mean overlap versus energy per spin. Inset zooms on the critical region. (\subref{fig:energy-T})~Energy per spin versus $T$. The monotonic relation justifies the energy axis in panel~(\subref{fig:cluster-mean-E}).}
  \label{fig:cluster-mean}
\end{figure*}

Figure~\ref{fig:cluster-mean-T} shows the mean cluster overlap $U_2^{(\mathrm{W})}$ versus temperature. It is finite in the ordered phase, drops sharply at $T_c$, and vanishes above it (two finite clusters rarely intersect), with the drop steepening with $L$. The overlap does not diverge anywhere: at $T_c$ it passes smoothly through a finite value, so although it behaves like an order parameter, it is a bounded, non-singular quantity, and the size dependence at $T_c$ discussed below is a slow power-law decay of that finite value rather than any critical divergence. The ordered-phase plateau sits at $\approx0.44$ rather than unity because only the $2N/3$ active spins can join a cluster. This value is exact: as $T\to0$ each cluster fills both active sublattices, i.e. the complement of the frozen sublattice, so two successive clusters overlap in $2/3$ of the lattice when the same sublattice is frozen (probability $1/3$) and $1/3$ otherwise (probability $2/3$), giving
\begin{equation}
\begin{aligned}
\langle U_n^{(\mathrm{W})}\rangle\big|_{T\to0} &{=} \tfrac13\cdot\tfrac23+\tfrac23\cdot\tfrac13 {=} \tfrac49, \\[4pt]
\mathrm{Var}\big|_{T\to0} &{=} \tfrac29-\big(\tfrac49\big)^2 {=} \tfrac{2}{81},
\end{aligned}
\label{eq:plateau}
\end{equation}
both matching the data. The residual $T{=}0$ variance is purely algorithmic, set by the random sublattice choice rather than by spin fluctuations.

Against energy per spin (Fig.~\ref{fig:cluster-mean-E}), via the monotonic $\epsilon(T)$ relation of Fig.~\ref{fig:energy-T}, the curves collapse away from criticality and separate only near $\epsilon_c$ (inset).

\begin{figure}[!htb]
  \centering
  \includegraphics[width=0.85\columnwidth]{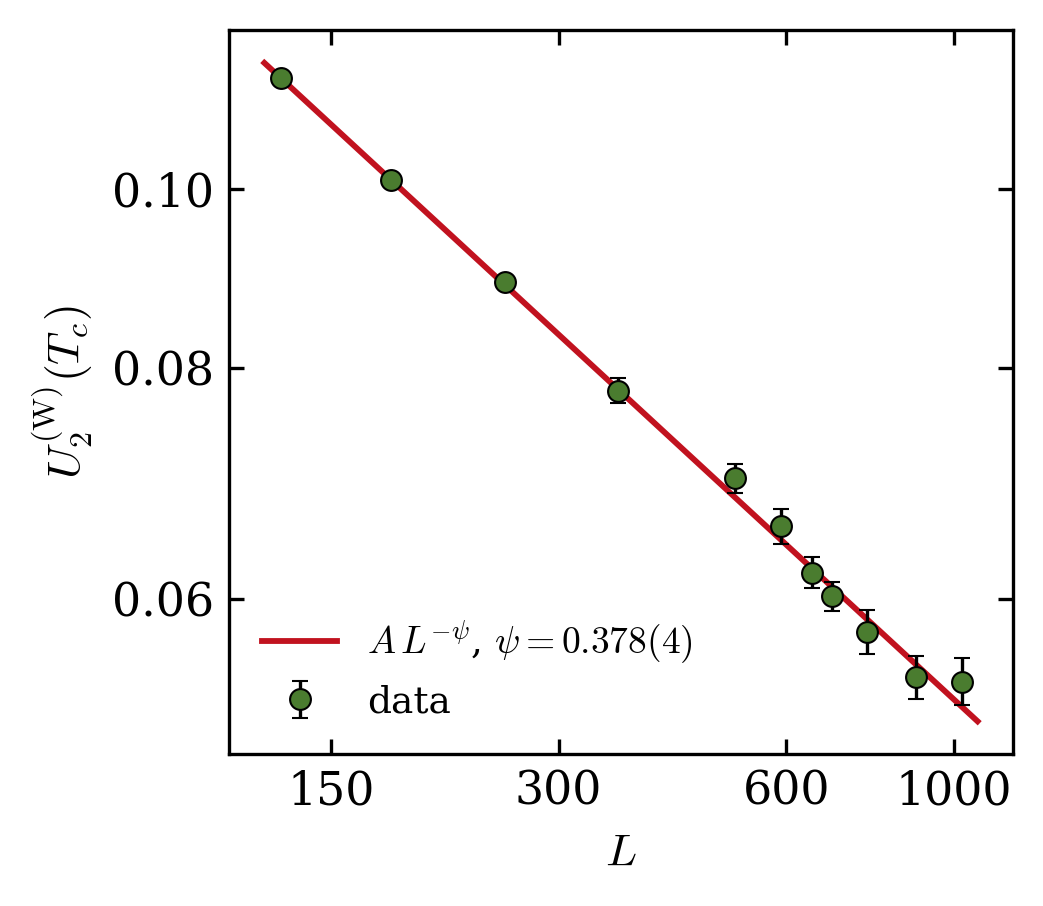}
  \caption{Finite-size scaling of $U_2^{(\mathrm{W})}(T_c)$ versus $L$ (log--log) for eleven sizes $L{=}129$--$1023$, with jackknife errors from $\sim\!400$ independent replicas; the line is a pure power law $A\,L^{-\psi}$ with $\psi^{(\mathrm{W})}_{\mathrm{BW}}{=}0.378(4)$ (reduced $\chi^2\approx0.6$), more moderate than the Ising/Potts value.}
  \label{fig:cluster-fss}
\end{figure}

Finite-size scaling at $T_c$ (Fig.~\ref{fig:cluster-fss}) gives a clean power law $U_2^{(\mathrm{W})}(T_c)\sim L^{-\psi^{(\mathrm{W})}}$ with
\begin{equation}
\psi^{(\mathrm{W})}_{\mathrm{BW}} {=} 0.378(4),
\label{eq:psi_BW}
\end{equation}
the central result, obtained from a two-parameter fit to all eleven sizes. The exponent is well determined, and it is smaller than the value $\approx0.42$ reported for the Ising and $q$-state Potts models under standard Fortuin--Kasteleyn cluster dynamics~\cite{pile2026overlaps}. Essentially, Baxter--Wu clusters have no Fortuin--Kasteleyn counterpart, because the three-spin interaction has no bond representation and the decomposition depends on which sublattice is frozen. Thus the exponent cannot be inherited from a static cluster geometry. Instead it reflects the Novotny--Evertz freezing dynamics: the overlap of successive single clusters decays more slowly with system size when the clusters grow on the frozen honeycomb sublattice than when they are ordinary Fortuin--Kasteleyn clusters. The exponent also does not track any static order-parameter ratio: $\beta/\nu{=}1/8$ is shared by the Baxter--Wu and four-state Potts classes, yet the algorithmic exponent differs from the FK value, confirming that $\psi^{(\mathrm{W})}$ is set by the cluster-construction rule rather than by the equilibrium critical exponents.

\subsection{Overlap fluctuations}

\begin{figure*}[!t]
  \centering
  \begin{subfigure}{0.48\textwidth}
    \centering\includegraphics[width=\linewidth]{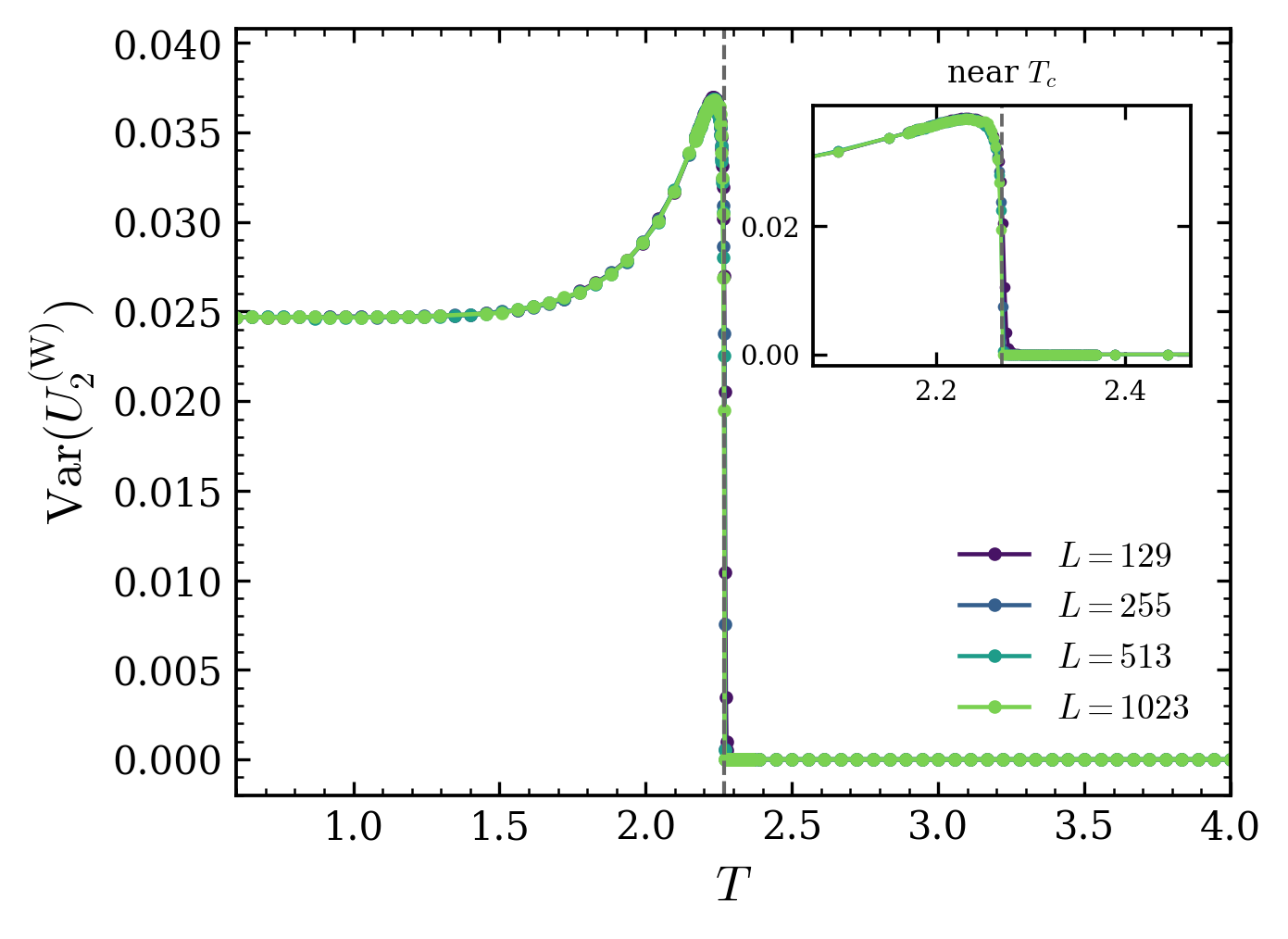}
    \caption{}\label{fig:cluster-var-T}
  \end{subfigure}\hfill
  \begin{subfigure}{0.48\textwidth}
    \centering\includegraphics[width=\linewidth]{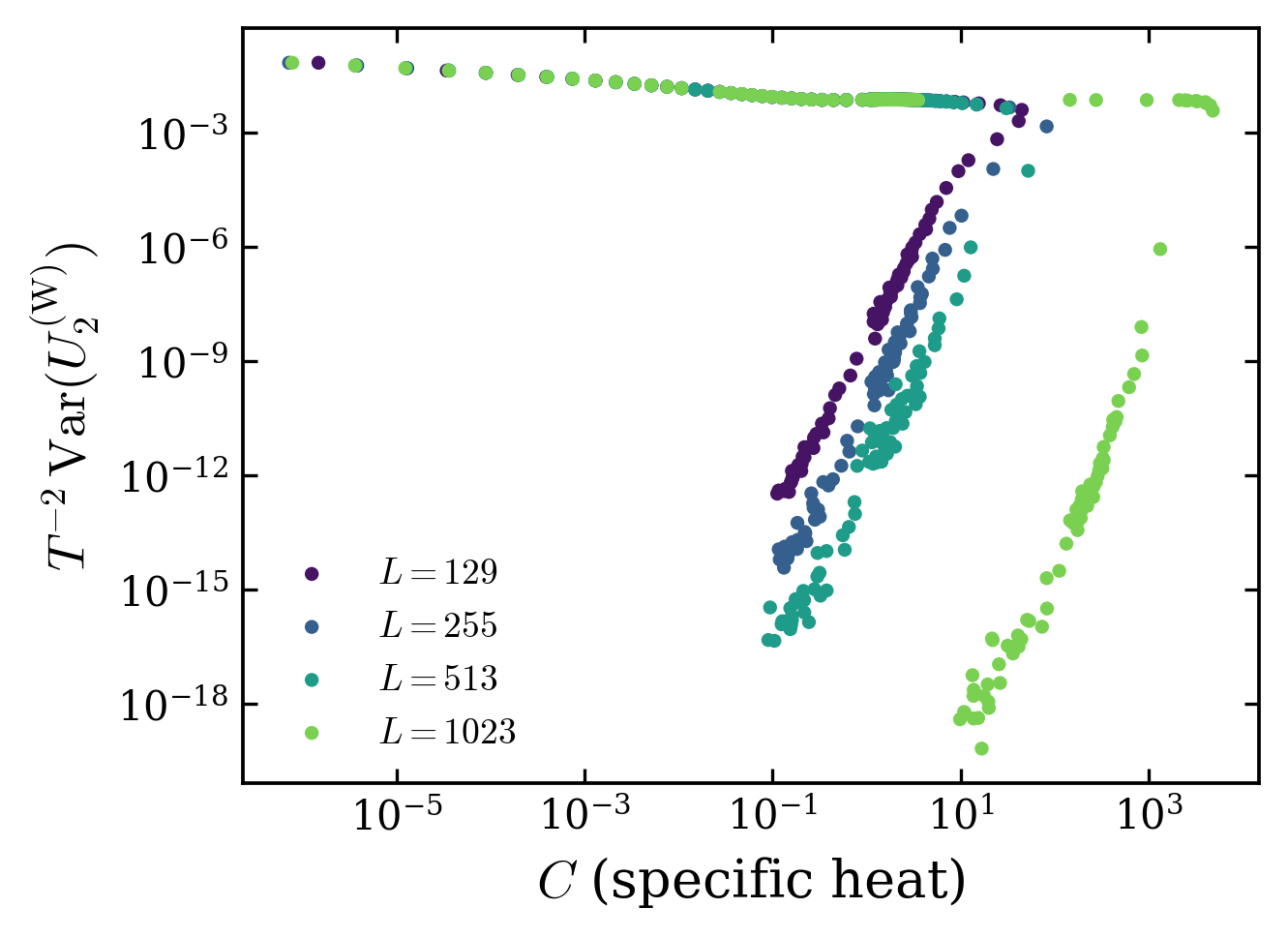}
    \caption{}\label{fig:cluster-var-C}
  \end{subfigure}
  \caption{(\subref{fig:cluster-var-T})~Variance $\mathrm{Var}(U_2^{(\mathrm{W})})$ versus $T$; inset on the critical peak. The peak sharpens with $L$; the low-$T$ plateau is $2/81$. (\subref{fig:cluster-var-C})~Rescaled variance $T^{-2}\mathrm{Var}(U_2^{(\mathrm{W})})$ versus $C$ (log--log). Upper branch $T<T_c$, lower branch $T>T_c$; they meet at the specific-heat peak.}
  \label{fig:cluster-var}
\end{figure*}

The variance $\mathrm{Var}(U_2^{(\mathrm{W})})$ (Fig.~\ref{fig:cluster-var-T}) develops a peak just below $T_c$ that narrows as $L$ grows, tracking the sharpening of the transition. The peak \emph{height}, however, does not grow with system size: it saturates at $\mathrm{Var}(U_2^{(\mathrm{W})})\big|_{\mathrm{peak}}\approx0.037$, essentially independent of $L$ over the whole range $L{=}129$--$1023$. The fluctuations of the algorithmic overlap therefore stay bounded through criticality; the peak marks the transition by its position and width, not by a diverging amplitude. This is the fluctuation counterpart of the non-divergence of the mean, and it distinguishes $U_2^{(\mathrm{W})}$ from a genuine thermodynamic response function such as the magnetic susceptibility, whose peak grows as a power of $L$.

Away from the transition the variance settles onto its low-temperature plateau $\mathrm{Var}(U_2^{(\mathrm{W})})\to2/81$ [Eq.~(\ref{eq:plateau})]. This residual value is purely algorithmic: at low $T$ the spin configuration is essentially frozen, so all fluctuation in the overlap comes from the random choice of which sublattice is frozen at each step, not from thermal disorder.

Following Ref.~\cite{lev2019}, we plot the rescaled variance $T^{-2}\mathrm{Var}(U_2^{(\mathrm{W})})$ against the specific heat $C$ (Fig.~\ref{fig:cluster-var-C}). The data split into two branches---ordered ($T<T_c$, upper) and disordered ($T>T_c$, lower)---because $C$ is two-valued in $T$ on either side of the transition; the branches meet at the specific-heat peak, which locates $T_c$ without prior knowledge of the critical temperature. As $L$ increases the meeting point moves to larger $C$, following the growth of the specific-heat maximum expected for the four-state Potts universality class, while the overlap variance at that point stays finite---again consistent with a bounded, non-diverging algorithmic observable.

\subsection{Order-parameter view and multistep overlaps}

\begin{figure}[!htb]
  \centering
  \includegraphics[width=0.85\columnwidth]{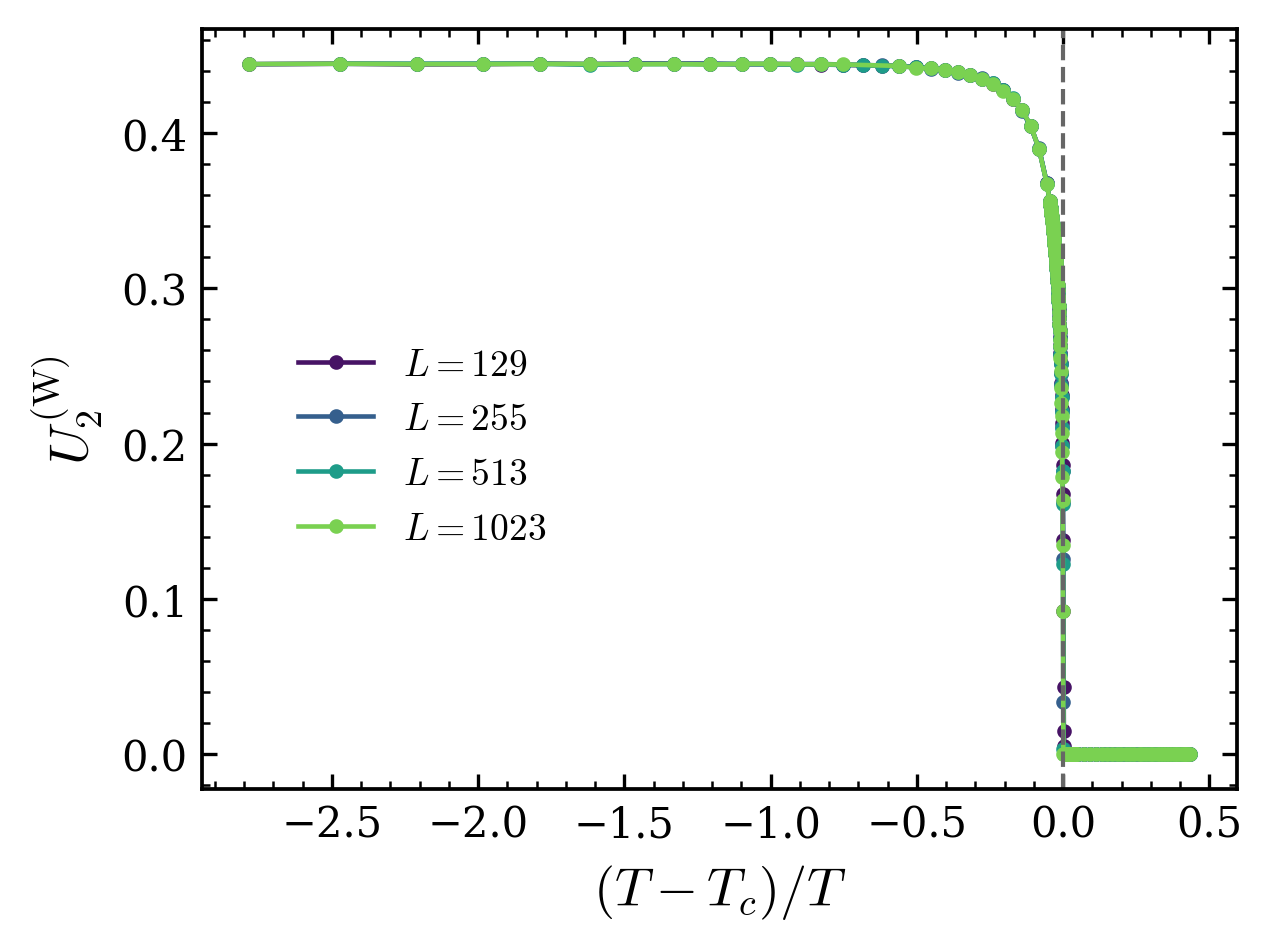}
  \caption{$U_2^{(\mathrm{W})}$ versus $(T-T_c)/T$; the overlap behaves like an order parameter, dropping from a finite plateau to zero without diverging.}
  \label{fig:cluster-reducedT}
\end{figure}

\begin{figure}[!htb]
  \centering
  \includegraphics[width=0.85\columnwidth]{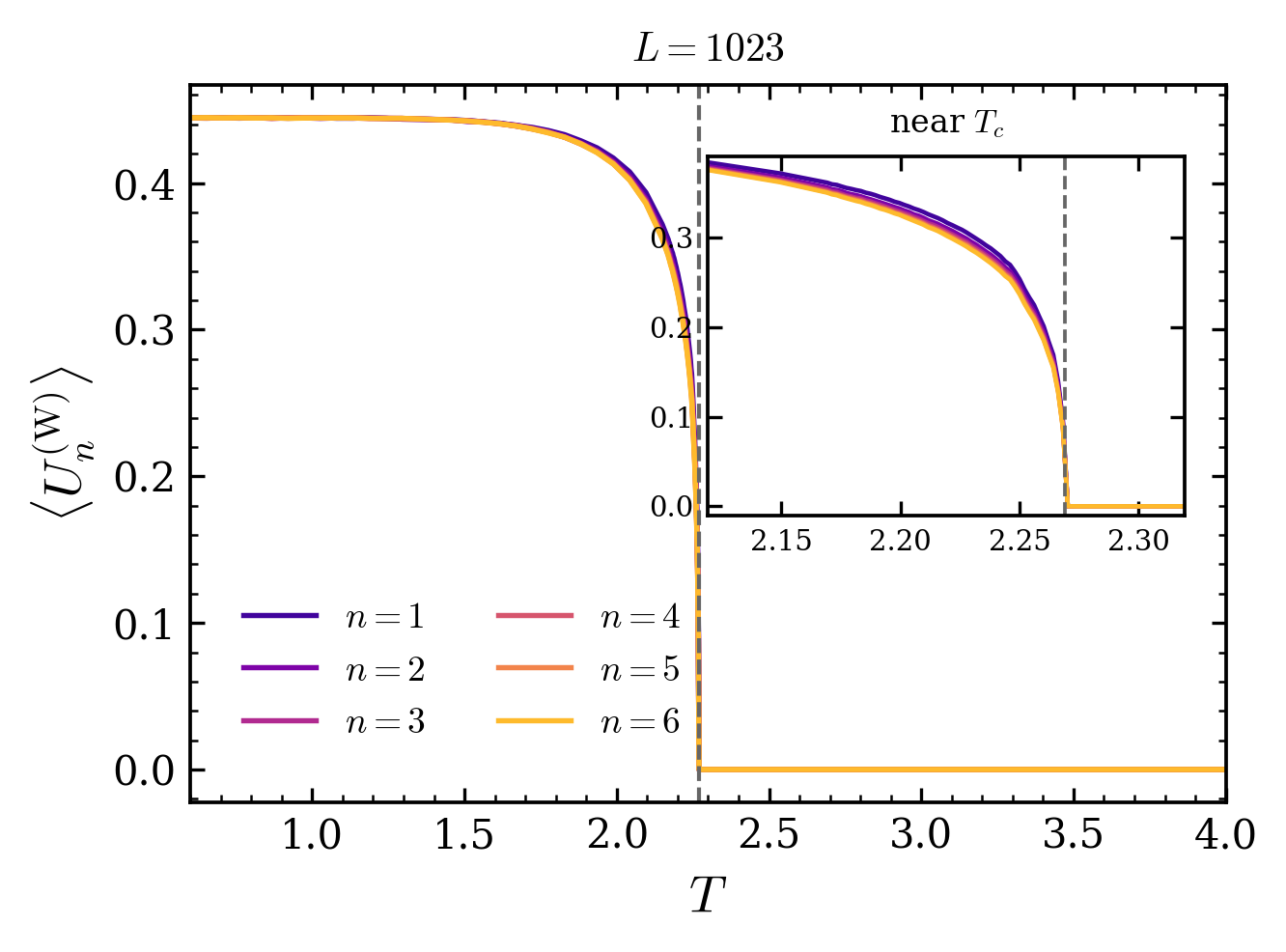}
  \caption{Multistep overlaps $\langle U_n^{(\mathrm{W})}\rangle$, $n{=}1$--$6$, $L{=}1023$ (inset: critical region); the structure persists for all $n$ and converges to a limiting profile.}
  \label{fig:cluster-multistep}
\end{figure}

Plotted against the reduced temperature $(T-T_c)/T$ (Fig.~\ref{fig:cluster-reducedT}), the overlap has the shape of an order parameter---finite below $T_c$ and zero above, while remaining bounded and non-divergent throughout. The multistep means $\langle U_n^{(\mathrm{W})}\rangle$, $n{=}1$--$6$ (Fig.~\ref{fig:cluster-multistep}), retain this shape and converge to a limiting profile as $n$ grows. Deep in the ordered phase this limit is already reached at $n{=}1$: all separations share the same plateau $\langle U_n^{(\mathrm{W})}\rangle\to4/9$ [Eq.~(\ref{eq:plateau})], independent of $n$ to four digits, because once the spin configuration is frozen the overlap depends only on the geometry of the frozen sublattices and not on how many update steps separate the two clusters. Near $T_c$ the curves for different $n$ separate slightly, since larger $n$ gives a marginally smaller overlap as successive clusters have more opportunity to migrate. The spread is small, and it closes again above $T_c$. The near $n$-independence of the profile shows that $U_2^{(\mathrm{W})}$ reflects a stationary property of the Markov chain rather than a short-time transient. We therefore use the single separation $n{=}2$ throughout.

\begin{figure}[!htb]
  \centering
  \includegraphics[width=0.85\columnwidth]{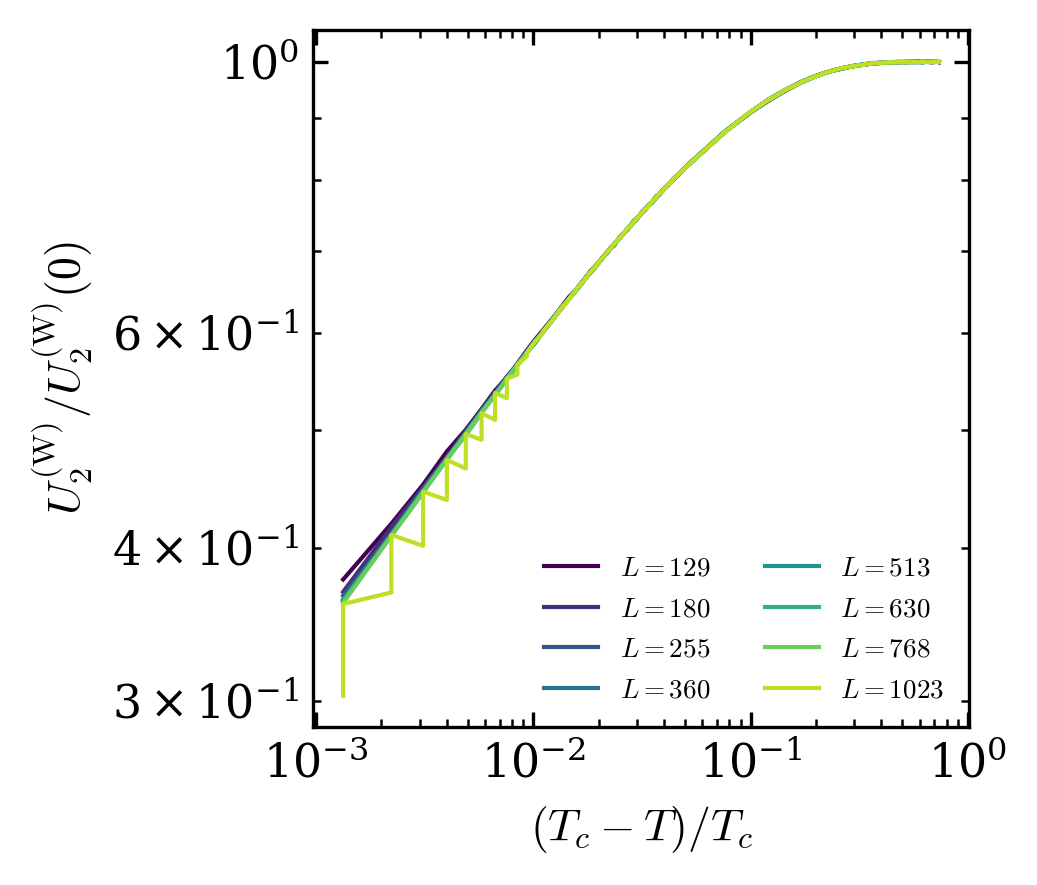}
  \caption{Log--log view of the ordered-phase Baxter--Wu overlap versus reduced temperature $(T_c-T)/T_c$, each size normalized to its low-temperature plateau; all sizes collapse onto a common profile that separates only near $T_c$.}
  \label{fig:bw-loglog}
\end{figure}

\begin{figure}[!htb]
  \centering
  \includegraphics[width=0.85\columnwidth]{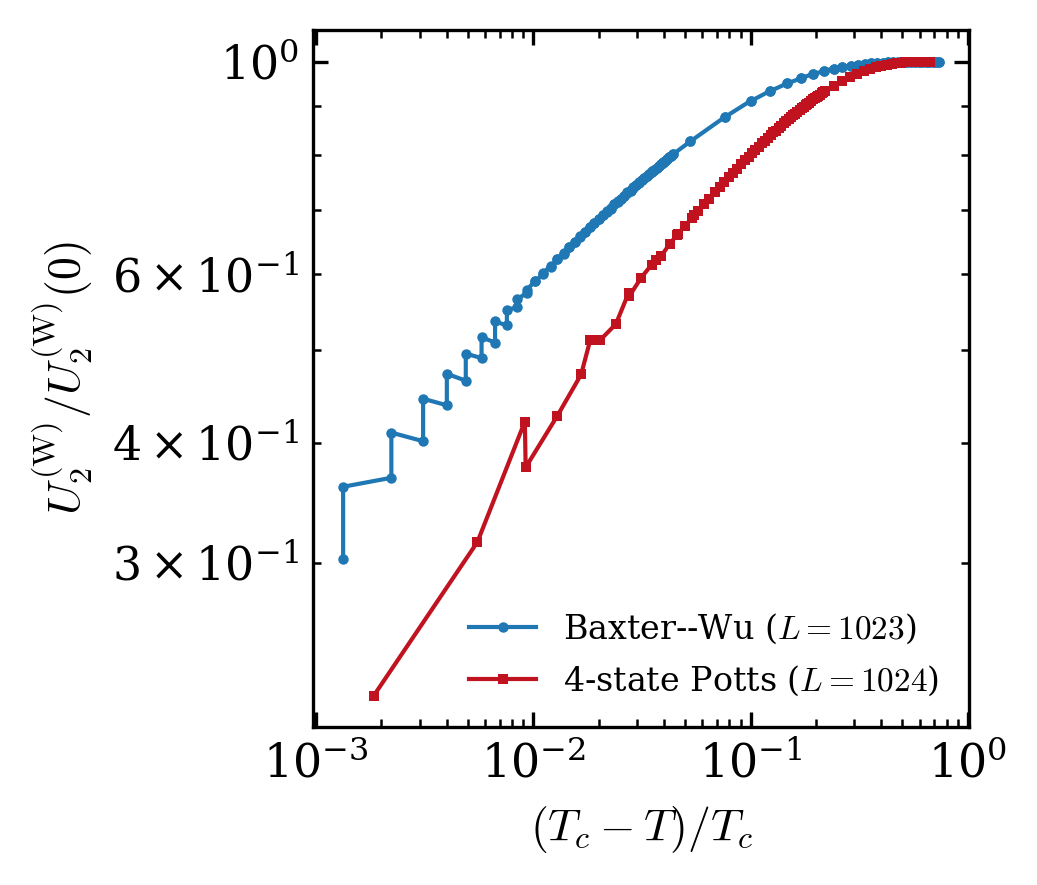}
  \caption{Log--log comparison of the normalized ordered-phase overlap for the Baxter--Wu ($L{=}1023$) and four-state Potts ($L{=}1024$) models versus $(T_c-T)/T_c$; the curves coincide deep in the ordered phase and separate near $T_c$, the Baxter--Wu overlap decaying less steeply.}
  \label{fig:bw-vs-potts-loglog}
\end{figure}

To compare the transition shape directly with the four-state Potts model, we normalize each overlap by its own ordered-phase plateau---$4/9$ for Baxter--Wu and unity for the standard Wolff updates of the four-state Potts model---so that both start at unity in the ordered phase. On a doubly logarithmic scale against the reduced temperature $(T_c-T)/T_c$ (Figs.~\ref{fig:bw-loglog} and~\ref{fig:bw-vs-potts-loglog}), the ordered-phase overlap of all Baxter--Wu sizes collapses onto a common profile that departs from the plateau only on approach to $T_c$, confirming that the shape is size-independent away from criticality. Comparing the two models on this scale, the curves in the ordered phase coincide and separate near $T_c$, with the Baxter--Wu overlap decaying more slowly. This difference comes not from the equilibrium critical behaviour but from the cluster-construction rule: only $2N/3$ of the spins are active under sublattice freezing, against all $N$ in standard Wolff.

\begin{figure}[!htb]
  \centering
  \includegraphics[width=0.85\columnwidth]{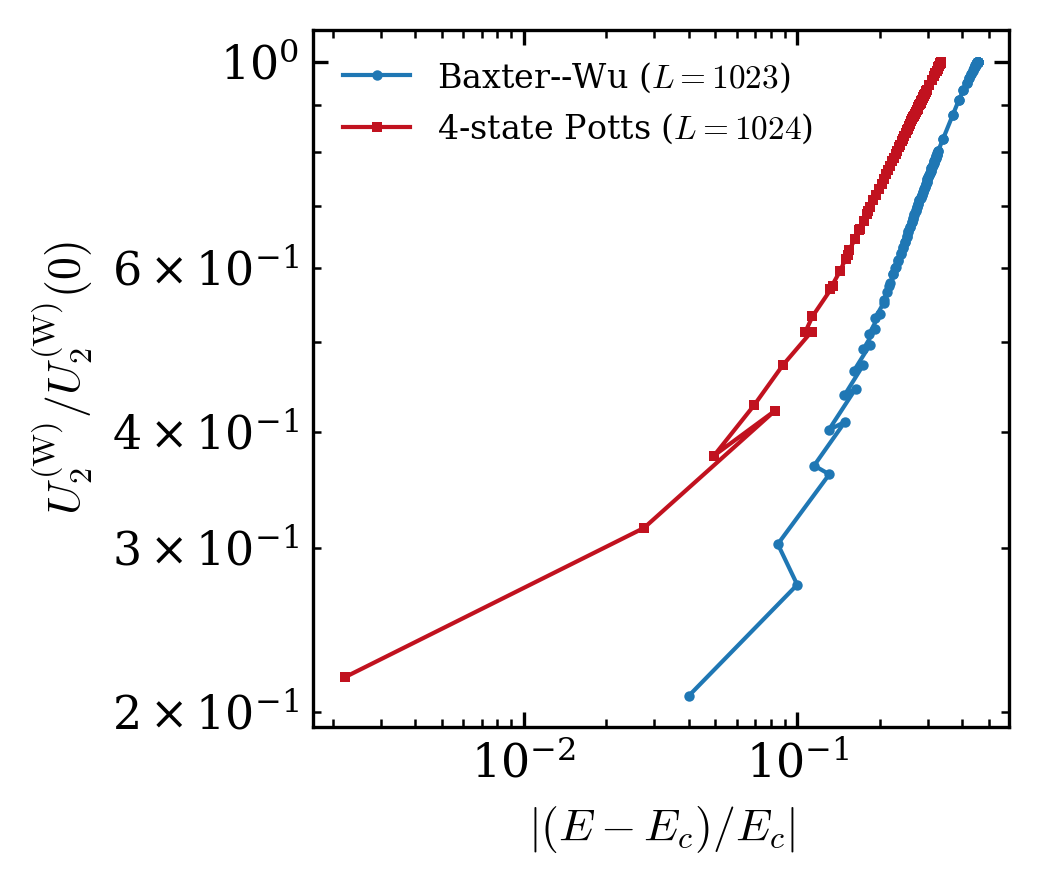}
  \caption{Single-cluster overlap versus reduced internal energy $|(E-E_c)/E_c|$ for the Baxter--Wu ($L{=}1023$) and four-state Potts ($L{=}1024$) models, each normalized to its low-temperature plateau; both axes are logarithmic (ordered phase, $E<E_c$). The critical energy corresponds to the left edge ($|(E-E_c)/E_c|\to0$). At equal reduced energy the Potts overlap lies above the Baxter--Wu one, the two converging only on approach to $E_c$.}
  \label{fig:bw-vs-potts-energy}
\end{figure}

The same contrast appears when the overlap is plotted against the reduced internal energy $|(E-E_c)/E_c|$ rather than temperature (Fig.~\ref{fig:bw-vs-potts-energy}), which removes the different temperature scales of the two models and compares them on a common thermodynamic axis. Both normalized overlaps fall from unity deep in the ordered phase ($E<E_c$) toward zero at the transition, but at any given reduced energy the four-state Potts overlap lies above the Baxter--Wu one---the Baxter--Wu overlap decays faster in energy---the two converging only as $E\to E_c$. The energy axis leads to the same conclusion as the temperature and finite-size data: the Novotny--Evertz overlap follows the transition with a weaker response than the Fortuin--Kasteleyn clusters of the Potts model.

\section{Discussion}\label{sec:discussion}

\begin{table}[!b]
\caption{Exact static exponents of the four-state Potts class (shared by Baxter--Wu) and the algorithmic single-cluster-overlap exponent measured here. The Baxter--Wu value is obtained under Novotny--Evertz sublattice-freezing dynamics; the Ising/Potts value~\cite{pile2026overlaps} is obtained under standard Fortuin--Kasteleyn cluster dynamics. The static exponents coincide across the two classes, but the algorithmic exponent does not.}
\label{tab:exponents}
\squeezetable
\begin{ruledtabular}
\begin{tabular}{lcc}
 & Baxter--Wu & $q{=}4$ Potts \\
\hline
$\alpha$ & $2/3$ & $2/3$ \\
$\beta$ & $1/12$ & $1/12$ \\
$\nu$ & $2/3$ & $2/3$ \\
$\beta/\nu$ & $1/8$ & $1/8$ \\
\hline
$\psi^{(\mathrm{W})}$ [FSS of $U_2(T_c)$] & $0.378(4)$ & $\approx0.42$~\cite{pile2026overlaps} \\
\end{tabular}
\end{ruledtabular}
\end{table}

The measured exponent $\psi^{(\mathrm{W})}_{\mathrm{BW}}{=}0.378(4)$ follows a clean power law across all eleven sizes and is smaller than the value $\approx0.42$ reported for the Ising and $q$-state Potts models under standard Fortuin--Kasteleyn dynamics. The static critical exponents of the Baxter--Wu model coincide with those of the four-state Potts class ($\alpha,\beta,\nu$ and $\beta/\nu$ in Table~\ref{tab:exponents}), so a shift in $\psi^{(\mathrm{W})}$ cannot come from the equilibrium universality class. Neither can it come from a static cluster geometry: the three-spin interaction has no bond representation, and the Novotny--Evertz clusters depend on which sublattice is frozen~\cite{vasilopoulos2026}, so there is no Fortuin--Kasteleyn object whose fractal dimension could set the exponent. The natural reading is therefore that $\psi^{(\mathrm{W})}$ is a property of the cluster-\emph{construction} rule. Under Novotny--Evertz freezing, each single cluster is grown on the honeycomb lattice of the two active sublattices---only $2N/3$ of the spins are eligible at any step, and successive updates freeze different sublattices---so the overlap of consecutive clusters decays more slowly with $L$ than for ordinary Fortuin--Kasteleyn clusters that can span the whole lattice. The result is that the algorithmic-overlap exponent is not universal across cluster algorithms: it takes a common value across Ising and Potts under FK dynamics, but a smaller value for the Baxter--Wu model under Novotny--Evertz dynamics. The overlap exponent therefore characterizes the update scheme rather than the equilibrium critical point.

\FloatBarrier

\section*{Data availability}

The data supporting this article are openly available at Ref.~\cite{data}.

\begin{acknowledgments}
This work is supported by the Russian Science Foundation (grant 25-11-00158).
\end{acknowledgments}

\section*{Author contributions}

\noindent\textbf{Ian Pil\'e:} Methodology, Software, Investigation, Writing -- original draft.\\
\textbf{Lev Shchur:} Conceptualization, Supervision, Writing -- review \& editing.

\end{document}